\documentclass[10pt]{article}

\usepackage{amsfonts}
\usepackage{amsmath}
\usepackage{latexsym}
\usepackage{amssymb}
\usepackage[mathcal]{euscript}
\usepackage{amscd}

\def\half{{1\over 2}}
\def\ben{\begin{equation}}
\def\een{\end{equation}}
\def\bea{\begin{eqnarray}}
\def\eea{\end{eqnarray}}
\input amssym.def
\input amssym.tex

\def \br{ {\bf r}}\def \bx{ {\bf x}}
  
 \def \half {\frac{1}{2}}

\def\bx{{\bf x}}

\begin{document}

\title{Dark Energy and the Schwarzian Derivative}

\author{
G. W. Gibbons\\
D.A.M.T.P.,\\  University of Cambridge, U.K.\\
and \\
Laboratoire de Math\'ematiques et de Physique \\
Th\'eorique, Universit\'e de Tours, France 
\\}

\date{\today}

\maketitle

\begin{abstract}
Theories with a time dependent  Newton's constant 
admit two  natural measures of time : 
atomic and astronomical. 
Temporal parametrisation  by 
$SL(2,{\Bbb R})$  transformations  gives rise to an equivalence
between theories  
with different time dependence's, including the special Case of no time
dependence, a fact noticed by   Mestschersky, Vinti and by Lynden-Bell.
I point out that theories with time dependent  dark energy densities  
admit three   natural measures of time : 
atomic and astronomical and de Sitter related by
temporal re-parametrizations and I 
extend   Mestschersky-Vinti-Lynden-Bell's result 
to cover this more general situation. 
I find a  consequent equivalence between theories in which the density
of dark energy is constant in time and in which it varies with time.
Strikingly a  time dependent cosmological constant 
changes by the addition of a Schwarzian derivative term
unless the temporal reparameterization belongs to $SL(2,{\Bbb R})$.
In General Relativity one may introduce a Schwarzian tensor 
to investigate how the notion of dark energy changes
under changes of conformal frame. The general theory is illustrated
in the case of Friedmann-Lemaitre metrics.

\end{abstract}

\section{Introduction} 
Until the development of Quartz, Ammonia and Caesium clocks,
time measurements were astronomical, and the default assumption was
that with respect to those  units,  Newton's law of gravity was     
independent of time.  The most economical assumption was 
then that the rate of  atomic processes are governed by the same
units \cite{Kelvin}. Thus the times which enter  Kepler's law and
Schr\"odinger's equation are the same and coincide with
those that enter Maxwell's equations. In which case   the  three `` fundamental 
constants
of physics'' $G,\hbar$ and $c$ would  indeed be constants.
and (Planck) units could be  adopted in which they  
be taken without loss of generality to equal unity \cite{Planck}.

However the constancy of all three constants has been questioned  
\footnote{A linguistic purist
might justifiably object to  this  oxymoron and even point out that
it only makes meaningful physical sense to say that dimensionless ratios of 
physical quantities may   vary with time. However 
since the construction of the requisite  dimensionless quantities 
is little more than an elementary undergraduate exercise I shall 
not trouble the reader by spelling it out in detail. Thus
for example  in what follows  ``time dependence  of Newton's constant'' 
should be understood in the same sense that one understands
the generally accepted usage of the phrase 
 ``time dependence  of Hubble's constant''.},  most
notably $G$ by, among others, Dirac \cite{Dirac}, Jordan \cite{Jordan} , 
Brans and Dicke \cite{Brans}.  
One may also question the constancy of $\hbar$ and $c$  
but the evidence against  any time variability  appears to be so strong
that in this paper I shall assume that they are indeed constant.
This and the evidence in favour  of the Weak Equivalence Principle  
is usually taken to justify ones belief in a spacetime
metric $g_{\mu \nu}$ such that propertime along the world line
of an idealised observer or experimenter coincides with 
atomic time and more generally the time of the standard model.
In principle, the other various constants of the standard model,
could vary with time, but current limits appear to be extremely stringent
and so  in this paper I shall assume that such things
as  the ``fine structure constant''\footnote{i.e. Sommerfelds's constant}
are indeed constant.
If this is not true, we would for example, have to introduce
Stoney time \cite{Stoney,Barrow1}.

While the  precision of  Einstein's equations  
have now been tested to an impressive level of accuracy,
limits on the time dependence  of Newton's constant $G$  
over cosmological times are not as stringent 
and  moreover the discovery of cosmic acceleration 
\cite{Riess:1998cb,Perlmutter:1998np}  quantified by an 
effective cosmological \footnote{Perhaps with more justice
nowadays this should be called  the de Sitter's  constant} constant $\Lambda$  
has suggested that one should question
the constancy in time of both.  This could arise , for example if both $G$
and $\Lambda$ 
depend locally on one or more scalar fields $\phi$  and  contribute  
(in which units  chosen so that $c=1$) to the action a term   

\ben
\frac{1}{16 \pi} \int  \Bigl (  \frac{R}{G(\phi)} - 2 \Lambda(\phi) \Bigr ) 
\sqrt{-g}  + \dots   \label{action}
\een
where the ellipsis denotes the contribution from the standard model
and the scalar $\phi$. By means  of a suitable Weyl  conformal rescaling  
\ben
g_{\mu \nu} \rightarrow \tilde  g_{\mu \nu} = \Omega^2 (\phi)  g_{\mu \nu} 
\label{Weyl} \een
the action becomes 
\ben
\frac{1}{16 \pi} \int  \Bigl (  \frac{\tilde R}{\tilde G(\phi)} - 
2 \tilde \Lambda(\phi) \Bigr ) \sqrt{-\tilde g}  + \dots   
\een
By choosing $\Omega(\phi)$ appropriately one may  arrange that
$\tilde G= {\rm constant}$, or $\tilde \Lambda = {\rm constant} $
but in general not both. It is customary to refer to the metric
$g_{\mu \nu}$ as the Jordan conformal frame and that  metric $\tilde g_{\mu \nu}$ 
for which 
$\tilde G= {\rm constant}$ as the Einstein conformal frame.
Therefore  it seems not unreasonable to refer to that  metric  $\tilde g_{\mu \nu}$  for which 
$\tilde \Lambda = {\rm constant}$  as the De-Sitter  conformal frame.
Since the  Einstein conformal frame metric may be said to 
measure  Planck units, one may also say that  De-Sitter  conformal frame  
measures  De-Sitter  units.

In this paper  I shall show that under temporal re-parameterisations,
or more generally Weyl conformal rescaling, the cosmological constant
changes by the addition of a term involving the Schwarzian 
derivative, or more generally the Schwarzian tensor of the transformation.
In more detail, the plan of the paper is as follows.
In section 2 I recall some elementary, but little noticed, facts 
about temporal reparametrisations in classical mechanics.
showing  how potentials change by the addition of a  
Schwarzian derivative term and I relate this to 
a conformal rescaling  of the assciated  higher dinensonal Bargmann
metric from which the equstions of motion may be obtained by means of a
null reduction.   In section 3 this general theory is 
applied to a Newtinain cosmlogical model. In particular
I show that  the cosmlogical constant or equivalently of dark energy 
changes by a Scwhwazian  derivative term 
under temporal reparametrisations.
In section 4 I turn to the General Relativistic theory
and the behaviour under conformal transformations
showing that  Schwarzian tensor enters.
In section 5 I apply these resuts to 
the Friedmann-Lemaitre cosmologies
and in particular to the   $\Lambda$ CDM model both at the
 Newtonian and General
Relativistic levels. I note a  connection bewteen the so-called 
cosmologiacl scalars including acceleration and jerk and the 
Schwarzian derivative.
Section 6 is a  short conclusion in which I suggest it would be interesting
to see whether or how  the Schwarzian derivative enters the transformation
fomulae for the energy momentum  tensor in 3+1 dimensional CFT's.  
 
\section{Temporal Re-parameterisations}          

We begin by placing  the old  results of  Mestschersky\cite{Mestschersky}  
Vinti \cite{Vinti} and Lynden-Bell
\cite{Lyn}  in a more general context and 
extend them to the case of a cosmological constant.
Our starting  point will be at the  Newtonian 
level. Later we will make contact with  Einstein's
covariant  viewpoint. 

Consider a system of $N$ point particles of mass $m_a$ and 
position vector $\bx _a$  moving  in  ${\Bbb E} ^3$ 
and governed by the action
\ben
\int  \Bigl \{ \sum _{1\le a \le N} \half m_a {\dot \bx}_a^2   
- U (\bx _a, t) \Bigr  \} dt  \label{system}\,,
\een
where $\dot{}$ signifies differentiation with respect to $t$.
If we make the replacements
\ben
t = f(\tilde t) \,,\qquad \bx_a = \sqrt{f^\prime} \, {\tilde \bx} _a    
\label{trans}\een
where $\,^\prime$  denotes differentiation with respect to $\tilde t$ 
we find, up to a boundary term, that the action becomes 
\ben
\int  \Bigl \{ \sum _{1\le a \le N} \half m_a
 { \tilde \bx    ^\prime _a }\, ^2   
- \tilde U ( {\tilde \bx} _a , \tilde t ) \Bigr  \} d \tilde t 
\een
where
\ben
\tilde U ({\tilde \bx}_a , \tilde t ) = f^\prime U( \sqrt{ f^\prime } 
\, {\tilde x}_a  , f(\tilde t) )  + \frac{1}{4} \{f,\tilde t \}
 \sum _{1\le a \le N} m_a{\tilde x} _a ^2  
\een
and 
\ben
\{ f, \tilde t \} = \Bigl( \frac{f^{\prime \prime}} {f^\prime } \Bigr )
 ^\prime - \half  \Bigl( \frac{f^{\prime \prime}}  {f^\prime}  \Bigr ) ^2 = 
\frac{f^{\prime \prime \prime} } {f^\prime} - \frac{3}{2}
\Bigl( \frac{f^{\prime \prime} } {f^\prime }  \Bigr ) ^2  
\een
is the Schwarzian derivative \footnote{The notation
$ $ is due to Cayley \cite{Cayley} and should not be confused with that
for Poisson brackets. Schwarz originally used $\psi((f,t)$ , Klein
$[f]_t$ and Koebe $D(f)_t$ .   More recently
$S(f)$ has become common.}  of $f$ with respect to $\tilde t$.

It is illuminating to view  the transformation (\ref{trans}) 
as a diffeomorphism  acting on the  
Eisenhart lightlike lift 
\cite{Eisenhart1929,DuvalBurdetKunzlePerrin1985,
Duval,GibPat03,Minguzzi:2005kz,Minguzzi:2006wz,Minguzzi:2006gq,Minguzzi:2012cx} of the Lagrangian system (\ref{system}) to 
a system of null geodesics of the  Bargmann metric on  
${\Bbb E} ^{3N} \times {\Bbb R} ^2 $ given by 
\ben
ds_B ^2 = \sum_{1\le a \le N} m_a d {\bf x}_a ^2 + 2dt ds 
 + 2 U (\bx _a, t) dt ^2  \label{I} 
\een
with $t$ an affine parameter along the null geodesics
Null geodesics (but not their affine parametrisation) are 
well known to be independent of Weyl conformal rescaling of the metric and 
we note that substituting  (\ref{trans}) in (\ref{I}) 
gives 
\ben
ds^2_B = 
 f^\prime \Bigl\{ \sum_{1\le a \le N} m_a d {\tilde {\bf x}} _a ^2             + 2 d \tilde t d \tilde s  + 2\tilde U (\tilde x_a, \tilde t) d \tilde t ^2         \Bigr \} \label{II}
\een
where
\ben
\tilde s = s - \half  \frac{f^{\prime \prime}}{f ^\prime} 
\sum_{1\le a \le N} m_a {\tilde \bx }_a^2 \,. 
\een
The metric ({\ref{I}) and the metric
inside the brace in (\ref{II}) are therefore conformal  but  
are not in general isometric and so (\ref{trans}) is in general 
not a conformal isometry\footnote{An earlier  related observation on 
a time dependent oscillator, but with no reference
the Schwarzian derivative may be found in \cite{BDP}.}
However, if $U=0$ and 
\ben
\{f, \tilde t \} =0 \,, \qquad \Longleftrightarrow   \qquad f= \frac{A\tilde t + B}{C \tilde t  + D } \,, \quad AD-BC \ne 0 \,,
\een
we obtain the Moebius or fractional linear subgroup $PSL(2,{\Bbb R} )$ 
of proper conformal symmetries of the 13-dimensional 
non-relativistic conformal group of a system 
of free  non-relativistic  particles. 
In particular if $B=C=0$ , $f^\prime =A^2$ and (\ref{trans}) 
becomes 
\ben
t= A^2 \tilde t \,,\qquad \bx_a = A \tilde \bx_a  \,.  
\een  
Note that because $t$ and $\bx_a$ scale in different ways, 
non-relativistic   conformal symmetries do not preserve the speed of 
light and are quite distinct from relativistic conformal symmetries
as we shall see in detail in \S\ref{FriedmannLemaitre}.

\section{Newtonian Cosmology}

At the Newtonian level we choose \cite{Duval,GibPat03,Gibbons:2013msa}
\ben
U(\bx _a,t)  = - G(t) \sum _{1\le a <b \le N} \frac{m_a m_b}{|\bx_ a - \bx_b|} 
-\frac{\Lambda (t)}{6}  \sum_{1\le a\le N} m_a {\bx }_a ^2 \,.
\een
Thus 
\ben
\tilde U(\tilde \bx _a,\tilde t)  = - \tilde G(\tilde t)
 \sum _{1\le a <b \le N} \frac{m_a m_b}{|\tilde \bx_ a - \tilde \bx_b|} 
-\frac{\tilde \Lambda (\tilde t)}{6}  \sum_{1\le a\le N} m_a {\tilde \bx }_a ^2 \,,
\een
where
\bea 
\tilde G(\tilde t  )  &=&  \sqrt {f^\prime}  G(f(\tilde t) ) \label{G} \\ 
\tilde \Lambda (\tilde t) &=&  ( f^\prime ) ^2   \Lambda (f(\tilde t ) )
 - \frac{3}{2} \{ f,\tilde t \} \label{L} \,.
\eea

While temporal re-parameterisations keep us within the class
of models with a   time dependent  Newtonian attraction and 
time dependent cosmic repulsion,  if we start with
$\Lambda (t) =0$ , we shall find that in general 
a time dependent cosmological term  is induced. The 
exceptional case is, as explained in   \cite{Duval}, when   
\ben
\{f, \tilde t \} =0 \,, \qquad \Longleftrightarrow   
\qquad f= \frac{A\tilde t + B}{C + D \tilde t} \,, \quad AD-BC \ne 0 \,,
\een
and coincides with  that originally considered by 
 Mestschersky    \cite{Mestschersky}, Vinti \cite{Vinti}  and Lynden-Bell \cite{Lyn}
(see also Barrow \cite{Barrow2})  who noted that  
if  we assume that $G(t) = {\rm constant}$, we may  obtain  
an example of a time dependent   Newton's constant  
all of whose solutions may be obtained from  those  with 
a time-independent  Newton's constant.

We could ask whether starting from a case in which $\Lambda (t)  $ and $G(t)$ 
are constant we could obtain a case with $\tilde \Lambda =0$ .  
This would entail solving the equation
\ben
\frac{2}{3}  ( f^\prime ) ^2   \Lambda 
 =  \{ f,\tilde t \} =   \Bigl( \frac{f^{\prime \prime}} {f^\prime } \Bigr )
 ^\prime - \half  \Bigl( \frac{f^{\prime \prime}}  {f^\prime}  \Bigr ) ^2 = 
\frac{f^{\prime \prime \prime} } {f^\prime} - \frac{3}{2}
\Bigl( \frac{f^{\prime \prime} } {f^\prime }  \Bigr ) ^2    \,.
\een

Setting  
\ben
y= \frac{f^{\prime \prime}}{(f^\prime)^2} = - 
\half \bigl ( \frac{1}{f^\prime} \bigr )^\prime  
\,,
\een
we start  with
\ben
\frac{2}{3} \Lambda = y^\prime = \frac{3}{2} y^2 \,,
\een 
whence
\ben
\frac{3}{2 \Lambda} \frac{dy}{1- (\frac{3y}{2 \sqrt{\Lambda}} ) ^2} = 
d \tilde t  \,.
\een
Thus
\ben
y= \frac{2}{3} \sqrt{\Lambda} \tanh (\sqrt{\Lambda} ( \tilde t - \tilde t_0) )
=  \frac{2}{3} 
\Bigl ( \ln \bigl( \cosh (\sqrt{\Lambda} (\tilde t -\tilde t_0)  
\bigr ) \Bigr ) ^\prime  \,,
\een
and hence
\ben
\frac{2}{3} \ln \Bigl( \cosh \sqrt{\Lambda} (\tilde t-\tilde t_0) 
\Bigr )  = A - \frac{1}{2 (f^\prime)^2 } \,, 
\een
where $\tilde t_0$ and $A$ are constants of integration.

\subsection{CFT}
The formula for the change of the cosmological constant under
a temporal re-paramterisations may be written as 
\ben
- \frac{2}{3} \tilde \Lambda(\tilde t ) = - 
\frac{2}{3}  \Lambda( t ) \big( \frac{dt}{d \tilde t} \bigr )^2
   +\{ t,\tilde t \} \,. \label{quad1}
\een
or in terms of ``quadratic differentials''. 
\ben
- \frac{2}{3} \tilde \Lambda(\tilde t ) (d \tilde t)^2  = - 
\frac{2}{3}  \Lambda(t) (d t) ^2  +\{ t,\tilde t\}  (d \tilde t) ^2 \,.
\label{quad2} \een

 The asymmetry in (\ref{quad1}) and (\ref{quad2})
is ony apparent since \cite{Cayley})
\ben
\{t,\tilde t\} = - \{ \tilde t,t\} \bigl( \frac{d t}{d \tilde t} \bigr)^2 
\, ,\qquad \Longleftrightarrow \qquad  \{t,\tilde t\} (d \tilde t)^2 = 
- \{ \tilde t, t\} (dt )^2 \,.  
\een    
and so we have
\ben
 \Bigl( \frac{2}{3} \tilde \Lambda(\tilde t ) (d \tilde t)^2   -\half 
\{ t, \tilde t\} \Big  )  (d \tilde t) ^2   = \Bigl( 
\frac{2}{3}  \Lambda(t)   - \half \{ \tilde t, t\} \Bigl)  (d  t) ^2 \,.
\een

The similarity of (\ref{quad2}) to the formula for the transformation
of the energy momentum  tensor $T(z)$  under a holomorphic transformation
$z\rightarrow \tilde z=\tilde z(z)$ 
\ben
\tilde T(\tilde z) (d\tilde z)^2 = \Bigl(T(z) - \frac{c} {12} \{\tilde z, z\}  \Bigr ) (dz)^2  \,, \label{tranS} 
\een
will not be lost on readers familiar with two-dimensional
Euclidean Conformal Quantum Field Theories. Similar formulae
have arisen in the closely related context of calculations of the entropy
of branes \cite{Banados:1999tw}.

\section{Covariant formulation }\label{covariant} 

The concept of a Schwarzian  derivative, thought of as a quadratic differential
has  been generalised  by Osgood and Stowe 
in    real dimension greater 
than two to a Schwarzian Tensor  \cite{Osgood}.

If $\{M,g ,\phi\}$ is an $n$-dimensional (pseudo-)Riemannian manifold 
$\{M,g\}$ and $\phi$ a real valued function on $M$, then they define the 
Schwarzian  tensor $B_{\mu \nu}(\phi)$ 
of $\phi$ to be  the symmetric, trace-free second rank covariant 
tensor
\ben
_gB_{\mu \nu}(\phi)= \nabla _\mu \nabla _\nu \phi - \nabla _\mu \phi \nabla _\nu \phi - \frac{1}{n} \Bigl( \nabla ^2 \phi - (\nabla \phi)^2   \Bigr )g_{\mu \nu}  
\,.\een   
If $f: \{M,g\} \rightarrow \{\tilde M, \tilde g\}$ is a \emph{conformal 
transformation}, that is such that 
the pull-back of $\tilde g$ to $M$  is a \emph{Weyl conformal rescaling} of $g$ :
$f^\star \tilde g= e^{2 \phi} g$, then taking   $\phi=  \ln (||df||) $
 the \emph{Schwarzian derivative}  of $f$, written $Sf$, is defined to be
\ben
Sf_{\mu \nu} = {_gB}_{\mu \nu}(\phi) \,.
\een 
If  $n=2$ this reduces to the standard definition  because 
$g$ is the Euclidean metric and $z=x+iy$ , then
\ben
Sf_{\mu \nu} = B_{\mu \nu}(\ln |\frac{df}{dz}|   ) 
= \left ( \begin{array}{cc}  \Re Sf & -\Im Sf \\ -\Im Sf & -\Re f \\ \end{array}\right )_{\mu \nu}   \,, 
\een  
so that 
\ben
Sf_{\mu \nu} dx^\mu dx^\nu = \Re( Sf )(dx^2 -dy ^2) -2 \Im (Sf ) dx dy 
=\Re ( Sf dz ^2  ) \,.
\een
The Schwarzian tensor also behaves nicely under composition of 
conformal transformations: 
\ben
_gB_{\mu \nu} (\phi + \sigma) = {_gB}_{\mu \nu}(\phi) +{ _{\tilde g}B}_{\mu \nu}
(\sigma) \,.
\een
Moreover if  
\ben
\begin{CD} 
\{M,g\}  @>{h}>> \{\tilde M,\tilde g\} @>{f} >> \{{\tilde{\tilde M}},
{\tilde {\tilde g}}  \}     
\end{CD}
\een
then
\ben
S(f \circ h) = h^\star S(f) + S(h) \,. 
\een

For our purposes,  most important property of the Schwarzian tensor 
proved in \cite{Osgood} is that under a Weyl conformal rescaling
\ben
g_{\mu \nu} \rightarrow \tilde g_{\mu\nu} = e^{2\phi} g_{\mu \nu} 
\een
the trace-free Ricci tensor 
\ben
S_{\mu \nu} = R_{\mu \nu} - \frac{1}{n} R g_{\mu \nu}\een
changes by a shift of  a multiple of the   the  Schwarzian tensor
\ben
\tilde S_{\mu \nu}= S_{\mu \nu} -(n-2)  {\,_g}\negthinspace B_{\mu \nu}(\phi) \,.
\een 
Now suppose the trace free Ricci tensor $\tilde S_{\mu \nu}$  vanishes.
Then 
\ben
\tilde R_{\mu \nu} = \frac{1}{n} R \tilde g _{\mu \nu} 
\een
The Ricci identity $\tilde \nabla ^\mu \bigl( \tilde R_{\mu \nu}- \half
\tilde g_{\mu \nu} \tilde R \bigr )=0$ , implies that
\ben
\tilde R_{\mu \nu} = \Lambda  \tilde g{\mu \nu} \,, 
\een 
where \emph{$\Lambda $ is a constant}. 
Thus if one is able to  solve the equation
\ben
S_{\mu \nu}= (n-2)  {\,_g}\negthinspace B_{\mu \nu}(\phi) \,,
\een
for $\phi$, one may pass by a Weyl conformal rescaling to an Einstein metric
 metric for which
the density of dark energy is constant. Eliminating
$S_{\mu \nu}$ altogether is a rather strong condition
 (see \cite{Brinkmann1,Brinkmann2,Brinkmann3,Brinkmann4} for some results on this) and in what follows
we shall take a less restrictive but more explicit approach using 
Friedmann-Lemaitre metrics.  We conclude this section by
noting  {\it en passant} that there is a  relationship
between the Schwarzian tensor and  the family 
of e Bakry-\'Emery-Ricci tensors  recently introduced
into scalar-tensor theory by Woolgar \cite{Woolgar:2013yzk}.

\section{Friedmann-Lemaitre Models} \label{FriedmannLemaitre}

We now apply the theory above to the case of Friedmann and Lemaitre's
cosmic expansion.

\subsection{Newtonian treatment}
At the Newtonian level \cite{Duval,GibPat03,Gibbons:2013msa} 
we make the homothetic ansatz
\ben
\bx_a=a(t) \br _a 
\een
where $\br_a$ are independent of time, 
and constitute a central configuration \cite{BatGibSut03} and $a(t)$ is the scale factor satisfying 
the Raychaudhuri equation but now with time independent Newton's constant 
and  cosmological constant:
\begin{equation}  \label{Newtlamb51}
\frac{1}{a(t)} \frac{d^2 a(t)}{dt^2} = -\frac{G\tilde{M}}{a^3(t)} + \frac{%
\Lambda}{3}
\end{equation}
Note that (\ref{Newtlamb51}) is \emph{identical} to
the Raychaudhury equation resulting from the applying
the Einstein equations to a   Friedmann-Lemaitre spacetime  
containing just pressure-free matter and a constant cosmological term
$\Lambda$ (the so-called $\Lambda$CDM model\cite{Gibbons:2013msa}. 
Making the temporal reparameterisation (\ref{trans})
and the replacement 
\ben
a(t) = \sqrt{f^\prime} \tilde a
\label{scale} \een 
we obtain
\begin{equation}
\frac{1}{\tilde a(t)} \frac{d^2 \tilde a(t)}{d \tilde t^2} = 
-\frac{\tilde G\tilde{M}}{\tilde a^3(t)} + \frac{
\tilde \Lambda}{3}
\label{Newtlamb5}
\end{equation}
where $\tilde G$ and $ \tilde \Lambda$ are again given by  
(\ref{G}) and (\ref{L}) . 

\subsection{General Relativistic  treatment}
At the level of the Einstein equations
for  Friedmann-Lemaitre metrics we have
in Einstein conformal frame   
\ben
ds_E^2 = -dt ^2 + a^2(t) d \omega _k^2   \label{FL} 
\een
where $t$ is  Einstein  cosmic time, $a(t)$ the scale-factor
and $  d \omega _k^2$ the metric on a three- space of constant
curvature $k=1,0,-1$. If $\Lambda$ and $G$ are truly constant then 
The Einstein equations imply 
 \bea
\frac{\ddot a}{a} &=& - \frac{4 \pi G }{3} \bigl(\rho + 3P \bigr ) + 
\frac{\Lambda}{3}\label{Raychaudhuri}  \\  
\bigl (\frac{\dot a}{a} \bigr ) ^2 + \frac{k}{a^2} &=&   
\frac{8 \pi G } {3} \rho   + \frac{\Lambda}{3} 
\label{Friedmann} \\
\dot \rho + \frac{3 \dot a}{a} \bigl( \rho + P \bigr )    &=&0  \label{therm}
\eea
where $\rho$ is the energy density of the matter and
 $P$ its pressure. Of course if $P=0$ , as in the $\Lambda$CDM model, 
then  $\rho = \frac{\rho_0}{a^3} $ 
and we recover (\ref{Newtlamb51}) with $\tilde M= 4 \pi \rho_0$ .
Note that in general (\ref{Raychaudhuri},\ref{Friedmann}) and (\ref{therm}) 
are not independent: given any two,  the third follows as long
as $\dot a \ne 0$.

Since in the case of the the $\Lambda$CDM model 
the equations are the same, the formulae for the change
under the temporal reparamterization (\ref{trns}) are the same
as the Newtonian case.   

\subsection{Conformal Transformations} 
On the other hand,  
if we re-parameterise the cosmic time coordinate $t$ in (\ref{FL})
according to (\ref{trans}),  in 
order that we have a Weyl conformal rescaling as in (\ref{Weyl}) 
we need to make the replacement   
\ben
a(t) =  f^\prime \tilde a (\tilde t) 
\label{scaleW}\,. \een
As noted in \S 2 It is clear that  (\ref{scale}) and (\ref{scaleW}) 
not the same. In the one case we preserve the non-relativistic equations of motion
which in the Friedmann-Lemaitre metrics amount to
the Raychaudhuri equation  (\ref{Newtlamb51}). In the other,
we preserve the velocity of light.  In other words if we were to
pass from cosmic  time, which is usually taken to coincide
with atomic time, to another time, for example to  astronomical or 
to De-Sitter time, we should also have a variable speed of light theory.
     
Following  the discussion in \S\ref{Covariant}
we put 
\ben
ds_E^2 = -d\tilde t ^2 + \tilde a^2(t) d \omega _k^2   =   e^{2\phi} 
\Bigl \{ - d t^2 + a^2(t) d \omega _k^2        \Bigr \}      \label{WFL} 
\een
with
\ben
 a = e^{-\phi} \tilde a  \,,\qquad d t = e^{-\phi} d\tilde t 
\,.\label{WFL2}\een
Then 
\ben
\frac{\dot a}{a}= e^\phi  \Bigl (\frac{\tilde a^\prime}{\tilde a} 
 - \phi ^\prime \Bigr ) 
\een

Thus the Friedmann equation (\ref{Friedmann}) becomes
\ben
\Bigl (\frac{\tilde a^\prime}{\tilde a} 
 - \phi ^\prime \Bigr )^2 + \frac{k}{{\tilde a} ^2}
= \frac{8 \pi G}{3} e^{-2 \phi}\rho   + e^{-2 \phi} \frac{\Lambda}{3} 
\label{Friedmannprime}\een
and (\ref{therm}) becomes
\ben
\rho ^\prime + 3 (\rho + P)  \Bigl (\frac{\tilde a^\prime}{\tilde a} 
 - \phi ^\prime \Bigr )   =0\,. \label{thermprime}
\een

The last term in the modified Friedmann equation (\ref{Friedmannprime})
may be understood  from the identity
\ben
\Lambda \sqrt{-g} dt d^3x  = e^{-2\phi} \Lambda \sqrt{-\tilde g} d \tilde t d^3 x  \,.\een
It is how one might naively  expect the cosmological constant to transform
from the way it appears in the action (\ref{action}). 

Under (\ref{WFL2}) The Raychaudhuri equation (\ref{Raychaudhuri}) equation 
becomes 
\ben
\frac{\tilde a^{\prime \prime} }{\tilde a} 
- \frac{1}{\tilde a} \bigl(\tilde a \phi ^\prime  \bigr ) ^\prime
= -\frac{4 \pi G }{3}  e^{-2 \phi} \bigl(\rho +3 P \bigl ) \,. \label{Raychudhurimod}
\een

Thus we obtain
\ben
\frac{{\tilde a} ^{\prime \prime} }{\tilde a} = -\frac{4 \pi G }{3}   
 \bigl(\tilde \rho + 3\tilde P)  \label{Raychudhuriprime}
\een 
with
\ben
\frac{1}{\tilde a} \bigl(\tilde a \phi ^\prime  \bigr ) ^\prime 
-\frac{4 \pi G }{3}  e^{-2 \phi} \bigl( \rho  +3P\bigr ) = - 
\frac{4 \pi G }{3}         
\bigl(\tilde \rho  +3 \tilde P \bigr ) \label{phieqn} \,.\een

That is
\ben
\frac{1}{a} \frac{d}{dt}(  a \frac{d \phi}{dt}) - 
\frac{4 \pi G }{3}  \bigl( \rho  +3P\bigr ) = -\frac{4 \pi G }{3}         
e^{2 \phi} \bigl(\tilde \rho  +3 \tilde P \bigr )  \label{phieqn2} \,.
\een
  
In effect the  dark energy density deduced from the
cosmic acceleration using the metric $\tilde g$
would correspond to a cosmological term 
\ben
\tilde \Lambda = - 4 \pi G( \tilde \rho + 3 \tilde P) 
\een 
and so if one is able to solve the second order
non-linear differential equation  (\ref{phieqn2}) for $\phi$
given $ \tilde \rho + 3 \tilde P$,
one may may find a conformal frame for any choice of  $\tilde \Lambda$.

\subsection{Cosmological  Scalars } \label{cosmicscalars}  

In discussing Friedmann-Lemaitre metrics 
one often introduces various   
cosmological scalars which are invariant under rescaling
the scale factor $a(t)$ and the cosmic time, $t$ by  independent 
constant factors \cite{Dunajski:2008tg}. 
Among them are the  Hubble constant,  deceleration
and jerk given respectively by 
\ben
H=\frac{1}{a}\frac{d a}{d t}
\,, \qquad q=-{a}
\Big(\frac{d a}{d t}\Big)^{-2}\frac{d^2 a}{d t^2}\, \qquad
j=a^2
(\frac{d a}{d t}\Big)^{-3}\frac{d^3 a}{d t^3} \,. 
\een
All current cosmological observations are consistent
with the so-called $\Lambda{\rm CDM}$ model which is equivalent to the 
statement that  the jerk
of the universes is one. For a recent discussion of direct measurements of 
the jerk and references to earlier literature see 
\cite{Zhai:2013fxa,Zhai:2013fxaBochner:2013uda}. 

We note here the relationship between between the Schwarzian derivative
to these three  cosmological scalars
\ben
\{a, t \}= H^2 (j- \frac{3}{2} q^2) \,. 
\een   
The Schwarzian derivative $\{a, t \}$ is invariant
under a constant rescaling of the scale factor $a(t)$ but not of
the cosmic time. By contrast the quantity
\ben
\frac{1}{(\frac{da}{dt})^2} \{a, t \} = \frac{1}{a^2} (j- \frac{3}{2} q^2) 
\een 
is invariant under Moebius transformations (including constant rescaling) of
the cosmic time coordinate $t$.

\section{Conclusion} 
In this paper I have shown how the Schwarzian derivative 
enters the formula for the change of the density of dark energy
under temporal re-parameterisations and how the Schwarzian tensor
enters when considering conformal rescalings of the metric.
I have illustrated this by considering a $\Lambda$CDM cosmology.
It is striking that a similar behavour crops up in the
change of the stress tensor of a two-dimensional CFT
under conformal transformations.  This seems to hint at a deeper
connection between dark energy and CFT's 
in 3+1 spacetime dimensions. In this connection it would be interesting
to see whether or how the Schwarzian derivative enters the transformation
formulae for the stress tensor.

\section{Acknowledgements}

I am grateful to the 
{\it Laboratoire de Math\'ematiques et de Physique Th\'eorique de l'Universit\'e de Tours}  for hospitality, and the  {\it R\'egion Centre} for a \emph{
``Le Studium''} research professor\-ship while the work reported here
was carried out.  I would also like to thank Christian Duval and Peter Horvathy
for helpful comments.

\end{document}